\documentclass[twocolumn,preprintnumbers,superscriptaddress,landscape,nofootinbib]{revtex4}
\usepackage[utf8]{inputenc}
\usepackage{amsmath}
\usepackage{amsthm}
\usepackage{float}
\usepackage{braket}
\usepackage{graphicx}
\usepackage{url}
\usepackage[colorlinks=true, pdfstartview=FitV, linkcolor=red, citecolor=blue, urlcolor=blue]{hyperref}
\usepackage{slashed}
\usepackage[normalem]{ulem}

\graphicspath{{./figures/}}

\newcommand{\diff}{\mathrm{d}}

\newcommand{\p}{\partial}

\newcommand{\be}{\begin{equation}}      
\newcommand{\ee}{\end{equation}}      
\newcommand{\bea}{\begin{eqnarray}}      
\newcommand{\eea}{\end{eqnarray}}

\newcommand{\im}{\mathrm{i}}
\newcommand{\e}{\mathrm{e}}

\newcommand{\calE}{\mathcal{E}}

\newcommand{\calH}{\mathcal{H}}

\begin{document}

\title{Real-time dynamics of Chern-Simons fluctuations near a critical point}

\author{Kazuki Ikeda}
\email[]{kazuki.ikeda.gtATkyocera.jp}

\affiliation{Research Institute for Advanced Materials and Devices, Kyocera Corporation,\\
Soraku, Kyoto 619-0237, Japan}

\author{Dmitri E. Kharzeev}
\email[]{dmitri.kharzeevATstonybrook.edu}

\affiliation{Center for Nuclear Theory, Department of Physics and Astronomy, Stony Brook University, Stony Brook, New York 11794-3800, USA}
\affiliation{Department of Physics, Brookhaven National Laboratory, Upton, New York 11973-5000}
\affiliation{RIKEN-BNL Research Center, Brookhaven National Laboratory, Upton, New York 11973-5000}

\author{Yuta Kikuchi}
\email[]{ykikuchiATbnl.gov}
\affiliation{Department of Physics, Brookhaven National Laboratory, Upton, New York 11973-5000}


\bibliographystyle{unsrt}

\begin{abstract}
The real-time topological susceptibility is studied in $(1+1)$-dimensional massive Schwinger model with a $\theta$-term. We evaluate the real-time correlation function of electric field that represents the topological Chern-Pontryagin number density in $(1+1)$ dimensions. Near the parity-breaking critical point located at $\theta=\pi$ and fermion mass $m$ to coupling $g$ ratio of $m/g \approx 0.33$, we observe a sharp maximum in the topological susceptibility. We interpret this maximum in terms of the growth of critical fluctuations near the critical point, and draw analogies between the massive Schwinger model, QCD near the critical point, and ferroelectrics near the Curie point. 
\end{abstract}

\maketitle

\section{Introduction}

Schwinger model \cite{schwinger1962gauge} is quantum electrodynamics in $(1+1)$ space-time dimensions. For massless fermions, Schwinger model is analytically solvable and equivalent to the theory of a free massive boson field \cite{schwinger1962gauge,brown1963gauge,sommerfield1964definition,zumino1964charge,hagen1967current,lowenstein1971quantum,casher1974vacuum,coleman1975charge}; however the model with massive fermions presents a challenge for analytical methods and has a rich dynamics. 

Recently, quantum algorithms have emerged as an efficient (and potentially  superior) way to explore the dynamics of quantum field theories, including the Schwinger model ~\cite{wallraff2004strong,majer2007coupling,Jordan:2011ne,Jordan:2011ci,Zohar:2012ay,Zohar:2012xf,Banerjee:2012xg,Banerjee:2012pg,Wiese:2013uua,Wiese:2014rla,Jordan:2014tma,Garcia-Alvarez:2014uda,Marcos:2014lda,Bazavov:2015kka,Zohar:2015hwa,Mezzacapo:2015bra,Dalmonte:2016alw,Zohar:2016iic,Martinez:2016yna,Bermudez:2017yrq,gambetta2017building,krinner2018spontaneous,Macridin:2018gdw,Zache:2018jbt,Zhang:2018ufj,Klco:2018kyo,Klco:2018zqz,Lu:2018pjk,Klco:2019xro,Lamm:2018siq,Gustafson:2019mpk,Klco:2019evd,Alexandru:2019ozf,Alexandru:2019nsa,Mueller:2019qqj,Lamm:2019uyc,Magnifico:2019kyj,Chakraborty:2019,Kharzeev:2020kgc,Shaw:2020udc,sahinoslu2020hamiltonian,Paulson:2020zjd,Mathis:2020fuo,Ji:2020kjk,Raychowdhury:2019iki,Davoudi:2020yln,Dasgupta:2020itb,2019PhRvD..99a4503M}. Previously, we have addressed the real-time dynamics of vector current \cite{Kharzeev:2020kgc} (a $(1+1)$-dimensional analog of the chiral magnetic current \cite{kharzeev2006parity,fukushima2008chiral}) induced by the ``chiral quench" -- an abrupt change of the $\theta$-angle, or the chiral chemical potential. In this paper, we will explore the connection between the real-time topological fluctuations and criticality using Schwinger model as a testing ground.





\begin{figure}[H]
    \centering
    \includegraphics[width=0.8\hsize]{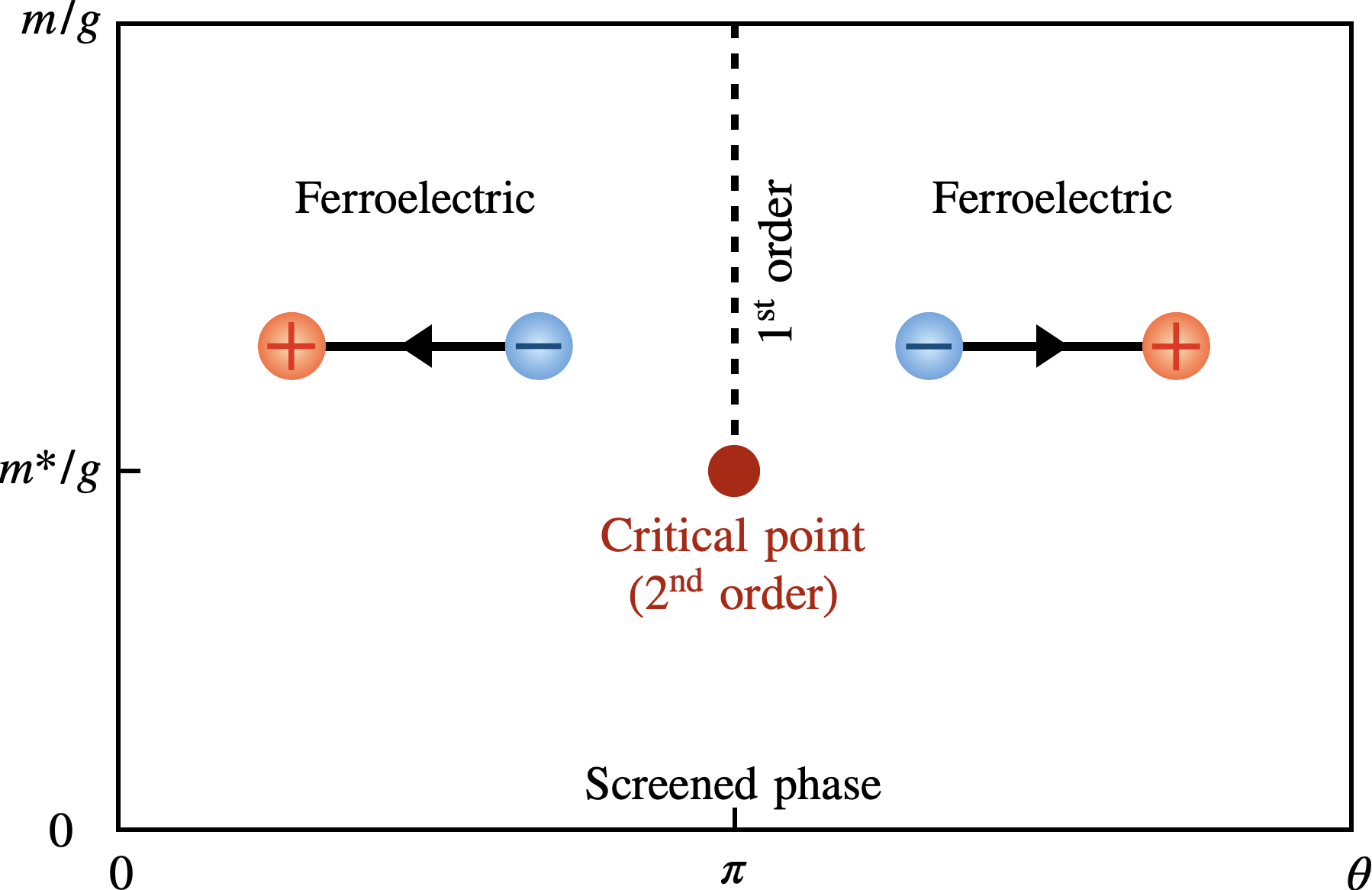}
    \caption{Phase diagram of the massive Schwinger model in the $(\theta,m/g)$ plane. At $\theta=\pi$ and large masses $m>m^\ast$, the ferroelectric phases with opposite orientations of electric field are separated by the line of the first order phase transition. This line terminates at $m^\ast\approx 0.33g$ at the critical point, where the phase transition is second order. 
    For small masses $m \ll m^\ast$, the electric field is screened by the production of light fermion-antifermion pairs.
    }
    \label{fig:phase_diagram}
\end{figure}

The massive Schwinger model possesses a quantum phase transition at $\theta=\pi$ between the phases with opposite orientations of the electric field, see Fig.\ref{fig:phase_diagram}. For $m\gg g$, this phase transition is first order. However, it was 
shown by Coleman \cite{coleman1976more} that the line of the first order phase transition terminates at some critical value $m^\ast$, where the phase transition is second order. The position of this critical point was established at $m^\ast\approx 0.33g$~\cite{schiller1983massive,hamer1982massive,byrnes2002density}; the resulting phase diagram is shown in Fig.\ref{fig:phase_diagram}. The phase diagram of the theory in the $(m/g, \theta)$ plane is thus reminiscent of the phase diagram of QCD in the $(T,\mu)$ plane of temperature $T$ and baryon chemical potential $\mu$ \cite{rajagopal2001condensed,stephanov2004qcd}.

To understand the physics behind this phase diagram, we need to recall the role of $\theta$-angle in the model. 
The action of the massive Schwinger model with $\theta$ term in $(1+1)$-dimensional Minkowski space is
\begin{align}
 S = \int\diff^2x\left[-\frac{1}{4} F^{\mu\nu} F_{\mu\nu} + \frac{g\theta}{4\pi}\epsilon^{\mu\nu}F_{\mu\nu} + \bar{\psi}(\im\slashed{D}-m)\psi\right],
\end{align}
with $\slashed{D}=\gamma^\mu(\p_\mu-\im gA_\mu)$.
Note that the gauge field $A_\mu$ and the coupling constant $g$ have mass dimensions 0 and 1, respectively.
Upon a chiral transformation, $\psi \to \e^{\im\gamma_5\theta}\psi$ and $\bar{\psi} \to \bar{\psi}\e^{\im\gamma_5\theta}$, the action is transformed to,\footnote{The action is invariant under this transformation only up to the boundary term.} 
\begin{align}
\label{action}
 S = \int\diff^2x\left[-\frac{1}{4} F^{\mu\nu} F_{\mu\nu}  + \bar{\psi}(\im\gamma^\mu D_\mu-m\e^{\im\gamma_5\theta})\psi\right].
\end{align}
It is clear from (\ref{action}) that the massive theory with a positive mass $m>0$ at $\theta=\pi$ is equivalent to the theory at $\theta=0$ but with a negative mass $-m$.  

\section{Topological fluctuations near the critical point}

Let us now discuss topological fluctuations in massive Schwinger model. For that purpose, it will be convenient to use
the Hamiltonian formalism with the temporal gauge, $A_0=0$.
From the action (\ref{action}) the canonical momentum conjugate to $A_1$ can be read off as $\Pi = \dot{A}_1$. The corresponding Hamiltonian is then given by
\begin{align}
\label{eq:H_Schwinger}
 H &= \int \diff x \left[\frac{1}{2}\Pi^2
 -\bar{\psi}(\im \gamma^1D_1 -m\e^{\im\gamma_5\theta})\psi\right],
\end{align}
with commutation relations $[A_1(x),\Pi(y)]=\im\delta(x-y)$, and $\{\psi(x),\bar{\psi}(y)\}=\gamma_0\delta(x-y)$.
\vskip0.3cm
In the bosonized description of the theory, the Hamiltonian is given by (see Appendix~\ref{app:theta} for details)
\begin{align}
\begin{split}
H = \int \diff x &\Big[ \frac{1}{2} \dot{\varphi}^2 + \frac{1}{2} (\partial_1{\varphi})^2 + \frac{\mu^2}{2} \left(\varphi + \frac{\theta}{2\sqrt{\pi}}\right)^2 \\
& - c  m  \mu \cos(2 \sqrt{\pi} \varphi) \Big]. 
\end{split}
\end{align}
The potential of the model is thus given by
\begin{align}
U(\varphi) = \frac{\mu^2}{2} \left(\varphi +\frac{\theta}{2\sqrt{\pi}}\right)^2 - c m \mu \cos{(2 \sqrt{\pi} \varphi)}, 
\end{align}
where $\mu = g/\sqrt{\pi}$ is the mass of the scalar boson, and the dimensionless coefficient $c$ is given by
$c = \e^\gamma/2\pi$
with the Euler constant $\gamma = 0.5774$.
The electric field is related to the boson field as
\be\label{el_bon}
E = -\mu \left(\varphi + \frac{\theta}{2\sqrt{\pi}}\right), 
\ee
where the first term is the quantum, dynamical contribution and the second one is the classical background induced by the $\theta$ angle, $E_\mathrm{cl} = -g\theta/2\pi$. 

The physics is periodic as a function of $\theta$ in the absence of boundaries -- as $\theta$ increases, the electric field becomes capable of producing a fermion-antifermion pair (or kink-antikink pair, in bosonic description) that screens it. 
At $\theta=\pi$ the potential takes the form
\be
U(\varphi) = \frac{\mu^2}{2} \left(\varphi+\frac{\sqrt{\pi}}{2}\right)^2 - c m \mu \cos{(2 \sqrt{\pi} \varphi)}.
\ee
When $m\gg \mu$, the potential has two well separated minima at $\varphi \approx 0,-\sqrt{\pi}$, associated with spontaneous symmetry breaking.
As follows from (\ref{el_bon}), the minima $\varphi = 0,-\sqrt{\pi}$ correspond to the electric fields $E =-g/2$ and $E=g/2$, respectively.

When $m \ll \mu$, there is a single minimum at $\varphi = -\sqrt{\pi}/2$, which according to (\ref{el_bon}) corresponds to the phase with no electric field. This is because at small $m$, the electric field is easily screened by the production of fermion-antifermion pairs - so we are dealing with the screened phase. At some critical value $m\approx m^\ast$, the effective potential becomes flat -- this corresponds to the critical point with a second order phase transition.
\vskip0.3cm
For $m > m^\ast$, the minima are separated by a potential barrier, and the transition between them (corresponding to the change in the direction of the electric dipole moment of the system) is first order -- for example, having a domain with an opposite orientation of the electric dipole moment would cost an additional energy due to the ``surface tension". Due to this extra energy, the fluctuations of the electric dipole moment at $m \gg m^\ast$ are suppressed. At the critical point, where $m \approx m^\ast$, the potential barrier between the two minima disappears, and the ``surface tension" of the domains with opposite orientations of the electric dipole moments vanishes. Because of this, the fluctuations of the electric dipole moment near the critical point are strongly enhanced. 
Here one can draw a useful analogy to the physics of ferroelectrics, where the electric susceptibility exhibits critical behavior near the Curie  point~\cite{rowley2014ferroelectric}. 

When the ratio $m/g$ becomes very small, the electric field is easily screened by the production of fermion-antifermion pairs (or kink-antikink pairs in the bosonized description), and  
the fluctuations of electric dipole moment again become small. Basing on this qualitative picture (that we will confirm below with a more formal treatment), we expect to see a maximum in the electric susceptibility near the critical point. We will show that this is indeed the case. 



\vspace{1em}

In order to characterize the topological fluctuation we compute the static topological susceptibility. It is the zero frequency and wavelength limit of the real-time two-point correlation function of the topological charge:
\begin{align}
\label{rate_def}
&\chi_\text{CS}
\nonumber\\
&=\lim_{\omega\to0}\lim_{k\to0}\int \diff^2 x\, \e^{\im\omega t-\im kx}\big[\langle Q(x) Q(0)\rangle -  \langle Q(x)\rangle\langle  Q(0)\rangle\big]
\nonumber\\
&=2\text{Re}\int \diff^2 x \Theta(t) \big[\langle Q(x) Q(0)\rangle -  \langle Q(x)\rangle\langle  Q(0)\rangle\big],
\end{align}
where $\Theta(t)$ is the Heaviside's step function.
$Q(x) = \partial_\mu K^\mu$ is the density of the Chern-Pontryagin number given by the divergence of the Chern-Simons current $K^\mu$. In $(1+1)$ dimensions, $K^\mu = (g/2\pi)  \epsilon^{\mu\nu} A_\nu$, and $Q= (g/4\pi) \epsilon^{\mu\nu} F_{\mu\nu}$, where $F_{\mu\nu}$ is the electromagnetic field strength tensor. The corresponding change of Chern-Simons number is $\Delta N_\mathrm{CS} = \int\diff^2x K^0 = (g/2\pi) \int\diff^2xA_1$; the Chern-Pontryagin number density is given by the electric field: $Q(x) = (g/2\pi)  E(x)$.
We call $\chi_\text{CS}$~\eqref{rate_def} the real-time topological susceptibility because it is computed from the real-time two-point function unlike the conventional topological susceptibility which is computed in the Euclidean space-time.
\vskip0.3cm

\section{Lattice Schwinger model}

\subsection{Lattice Hamiltonian}

For the purpose of numerical simulation, we place the theory~\eqref{eq:H_Schwinger} on a spatial lattice. We introduce the staggered fermion $\chi_n$ and $\chi^\dag_n$~\cite{Kogut:1974ag,Susskind:1976jm}, and lattice gauge field operators $L_n$ and $U_n$ with an integer $n$ labeling a lattice site; the lattice spacing is $a$. A two-component Dirac fermion $\psi=(\psi^1,\psi^2)^T$ is converted to a staggered fermion  $\psi^1(\psi^2)\to\chi_n/\sqrt{a}$ for odd (even) $n$. The gauge fields are replaced by $\e^{-\im agA_1}\to U_n$ and $\Pi\to -gL_n$, that are placed on a link between $n$th and $(n+1)$st sites. The resulting lattice Hamiltonian is,
\begin{align}
\label{eq:lattice_total_ham2}
 H &= \frac{a g^2}{2}\sum_{n=1}^{N-1} L_n^2
 -\frac{\im}{2a}\sum_{n=1}^{N-1}
 \big[\chi^\dag_{n+1} U_n\chi_{n}-\chi^\dag_{n}U_n^\dag\chi_{n+1}\big]
 \nonumber\\
 &+ m\cos\theta\sum_{n=1}^{N} (-1)^n \chi^\dag_n\chi_n
 \nonumber\\
 &+ \im\frac{m\sin\theta}{2}\sum_{n=1}^{N-1} (-1)^n \big[\chi^\dag_{n+1} U_n\chi_{n}-\chi^\dag_{n}U_n^\dag\chi_{n+1}\big],
\end{align}
with the Gauss law constraint:
\begin{equation}
L_{n+1}-L_n=\chi_n^\dag\chi_n- \frac{1-(-1)^n}{2}. 
\end{equation}
The right-hand side corresponds to the fermion density in terms of the Dirac fermions $\psi$. The second term reflects the fact that each component of Dirac fermion $\psi_1$ ($\psi_2$) is translated to a staggered fermion on odd (even) $n$.
We solve the constraint to eliminate the electric field operators,
\begin{equation}
\label{eq:solve_Gauss}
    L_n=\sum_{i=1}^n\left(\chi_i^\dag\chi_i - \frac{1-(-1)^i}{2}\right),
\end{equation}
where we have fixed the boundary electric field, $L_0=0$.
By enforcing the relation~\eqref{eq:solve_Gauss}, the states are automatically restricted to the physical ones.
We furthermore eliminate the link fields $U_n$ by the gauge transformation,
\begin{align}
    \chi_n\to g_n\chi_n,\quad
    \chi^\dag_n\to \chi^\dag_n g_n^\dag,\quad
    U_n \to g_{n+1}U_n g_n^\dag,
\end{align}
with
\begin{align}
    g_1=1, \quad g_n = \prod_{i=1}^{n-1}U^\dag_{i}.
\end{align}
In the present work, we limit $\theta$ to be $0$ or $\pi$ to study the critical behavior at $\theta=\pi$.
Upon absorbing $\cos\theta=\pm1$ in the second line of \eqref{eq:lattice_total_ham2} to the fermion mass $m$, we arrive at the Hamiltonian,
\begin{align}
\label{eq:lattice_total_ham3}
\begin{split}
 H &= \frac{a g^2}{2}\sum_{n=1}^{N-1} \left[\sum_{i=1}^n\left(\chi_i^\dag\chi_i - \frac{1-(-1)^i}{2}\right)\right]^2
 \\
 &-\frac{\im}{2a}\sum_{n=1}^{N-1}
 \big[\chi^\dag_{n+1}\chi_{n}-\chi^\dag_{n}\chi_{n+1}\big]
 \\
 &+ m\sum_{n=1}^{N} (-1)^n \chi^\dag_n\chi_n.
\end{split}
\end{align}
It is noted again that the massive theory with a positive mass $m>0$ at $\theta=\pi$ is equivalent to the theory at $\theta=0$ but with a negative mass $-m$. The Hamiltonian~\eqref{eq:lattice_total_ham3} accords with the latter viewpoint and will be used in what follows. Note that the method used here is not restricted to $\theta=0$ and $\pi$ but can be generalized to other values of $\theta$ without any difficulty. 


\subsection{Real-time topological susceptibility}

We compute the real-time topological susceptibility,
\begin{align}
&\chi_\text{CS}
=2\left(\frac{g}{2\pi}\right)^2\text{Re}\int \diff^2 x \Theta(t) \big[\langle E(x) E(0)\rangle -  \langle E(x)\rangle\langle  E(0)\rangle\big],
\end{align}
where $\Theta(t)$ is the Heaviside step function. Recall that the Chern-Pontyagin number density is proportional to the electric field in 1+1 dimensions~\eqref{rate_def}.
For the purpose of calculating the topological susceptibility we take the zero-wavelength limit followed by the zero-frequency limit,
\begin{align}
\begin{split}
 &\chi_\text{CS}
 = 2 \left(\frac{g}{2\pi}\right)^2V\text{Re}\int\diff t \Theta(t)(\langle\bar{E}(t)\bar{E}(0)\rangle-\langle\bar{E}(0)\rangle^2),
 \end{split}
\end{align}
where $V$ is spatial volume (length) and $\bar{E}(t):= \int\diff x E(t,x)/V$ is spatial average of electric field operator. We have used the translational invariance both in temporal and spatial directions, which approximately holds in a finite-size system.

On a lattice, it is given by
\begin{align}
 &\chi_\text{CS} =  2(N-1)\left(\frac{g}{2\pi}\right)^2ag^2
 \nonumber\\
 &\times \mathrm{Re}\int_0^T\diff t \Theta(t)(\langle\bar{L}(t)\bar{L}(0)\rangle
 -\langle\bar{L}(0)\rangle^2).
\end{align}
with $\bar{L}:=\frac{1}{N-1}\sum_{n=1}^{N-1}L_n$. Note that the temporal integral is also truncated by $T$ (not to be confused with temperature).
We numerically compute the dimensionless topological susceptibility,
\begin{equation}
\label{eq:diffusion}
 \frac{\chi_\text{CS}}{g^2} 
 = \frac{N-1}{\pi^2}\mathrm{Re}\int_0^{\hat{T}}\diff \hat{t} (t)(\langle\bar{L}(t)\bar{L}(0)\rangle
 -\langle\bar{L}(0)\rangle^2).
\end{equation}
with the dimensionless variables $\hat{t}:= (ag^2/2)t$ and $\hat{T}:= (ag^2/2)T$.

\section{Results and discussion}
\begin{figure*}
\begin{minipage}{0.329\hsize}
    \centering
    \includegraphics[width=\hsize]{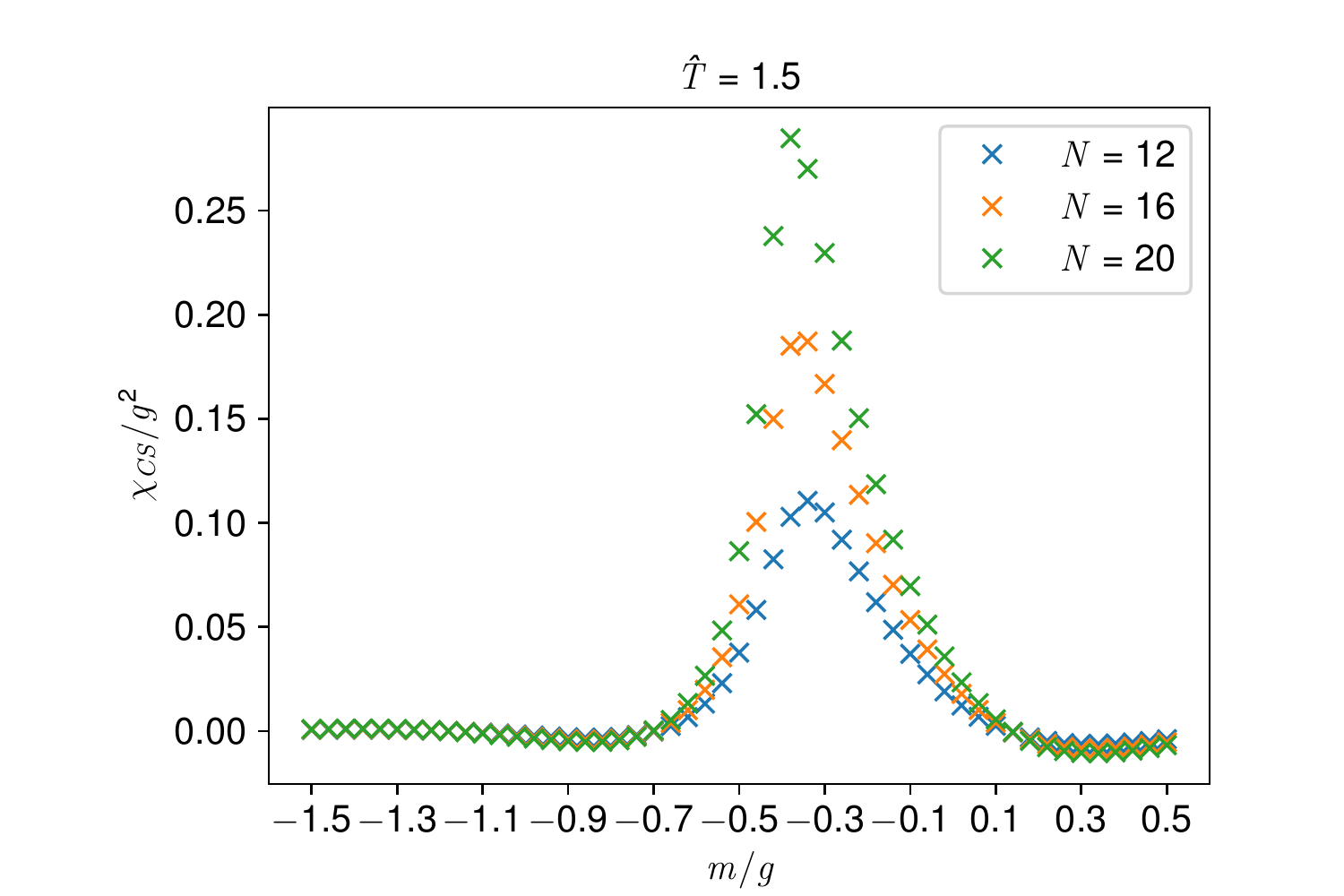}
\end{minipage}
\begin{minipage}{0.329\hsize}
    \centering
    \includegraphics[width=\hsize]{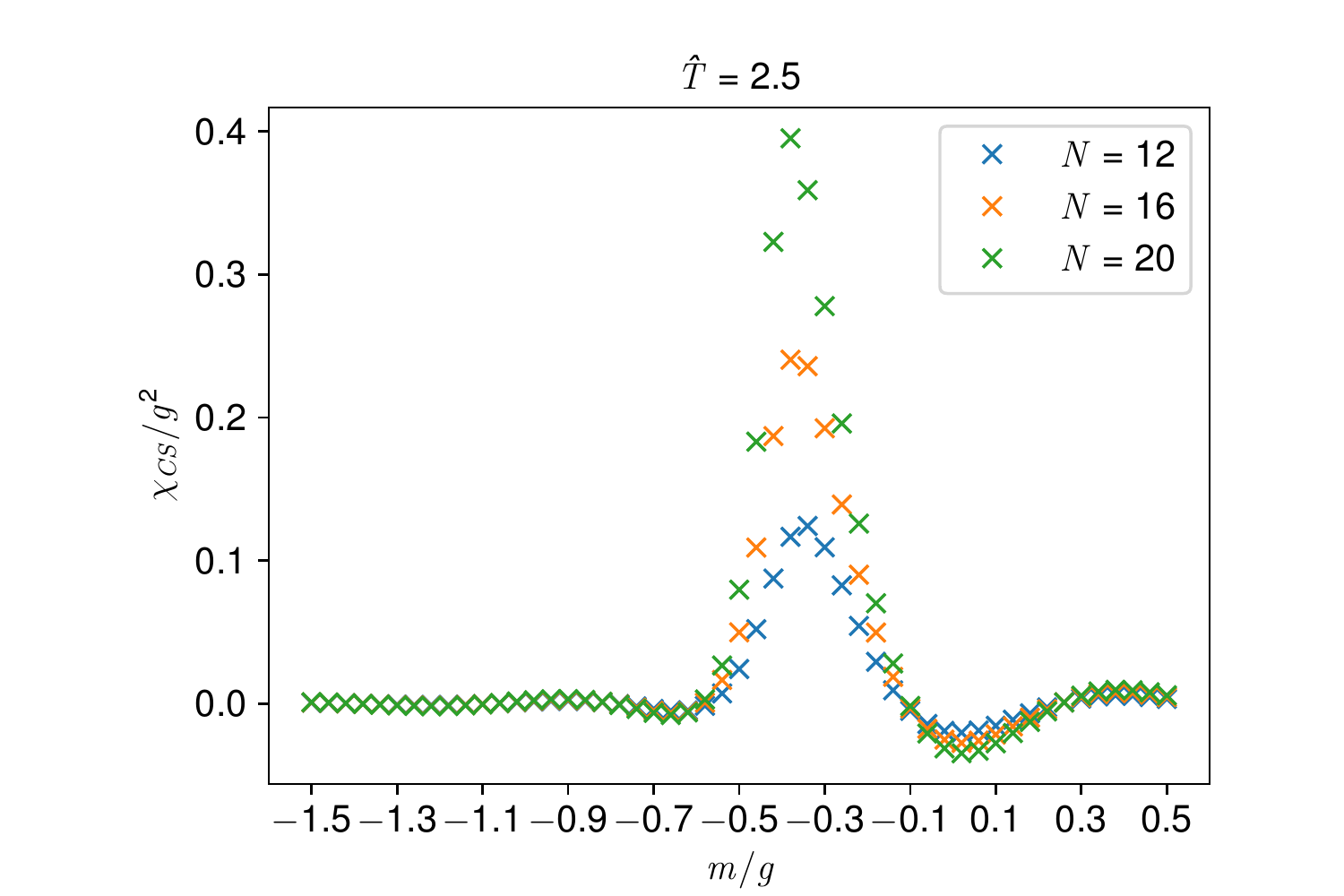}
\end{minipage}
\begin{minipage}{0.329\hsize}
    \centering
    \includegraphics[width=\hsize]{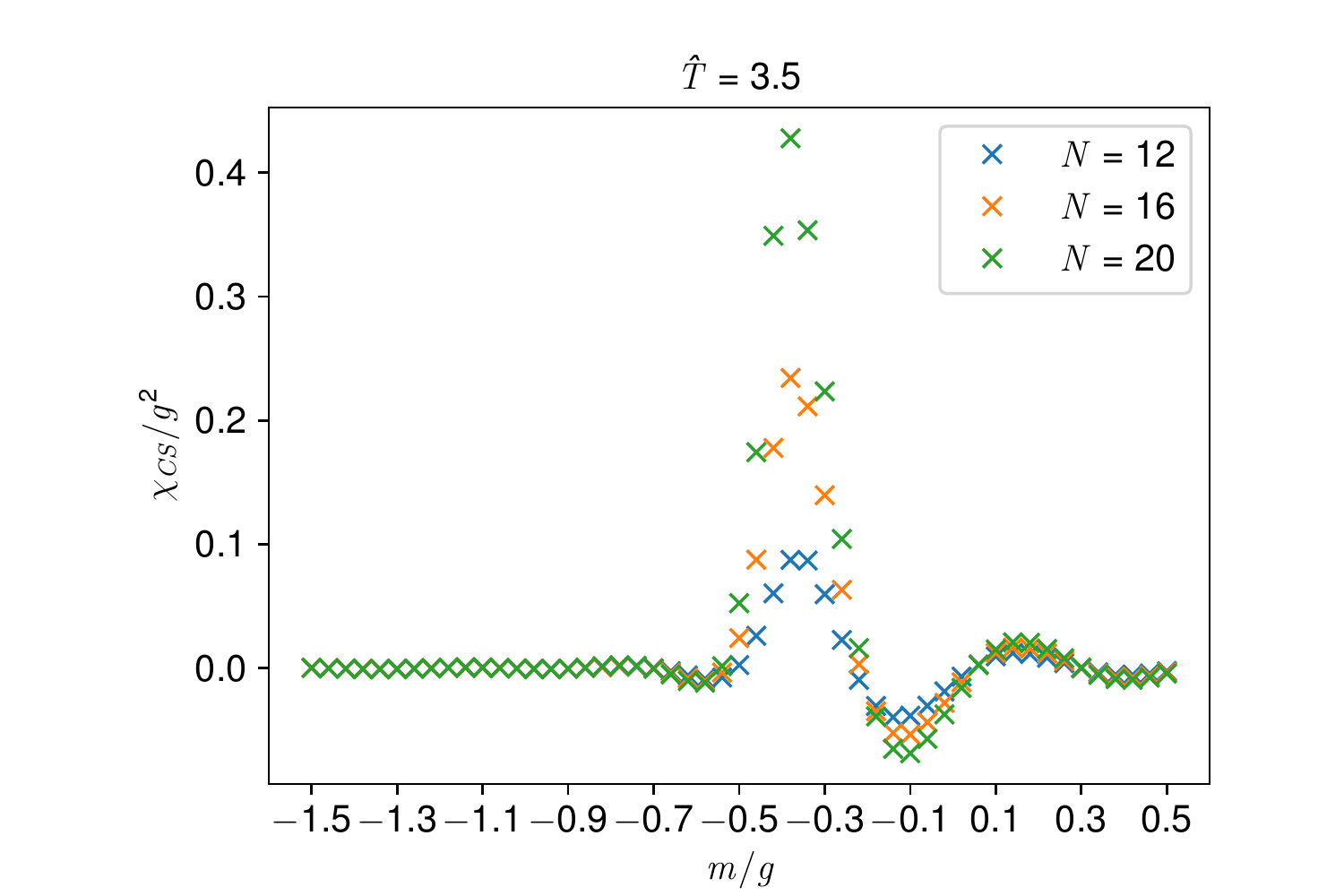}
\end{minipage}
    \caption{Dimensionless real-time topological susceptibility $\chi_\text{CS}/g^2$ as function of $m/g$ at different values of lattice size $N$ and range of temporal integral $\hat{T}=(ag^2/2)T$. The topological susceptibility is computed using the Hamiltonian~\eqref{eq:lattice_total_ham3}, where $\theta$ is limited to $0$ or $\pi$.
    }
    \label{fig:TS}
\end{figure*}

\begin{figure*}
    \centering
    \includegraphics[width=0.5\hsize]{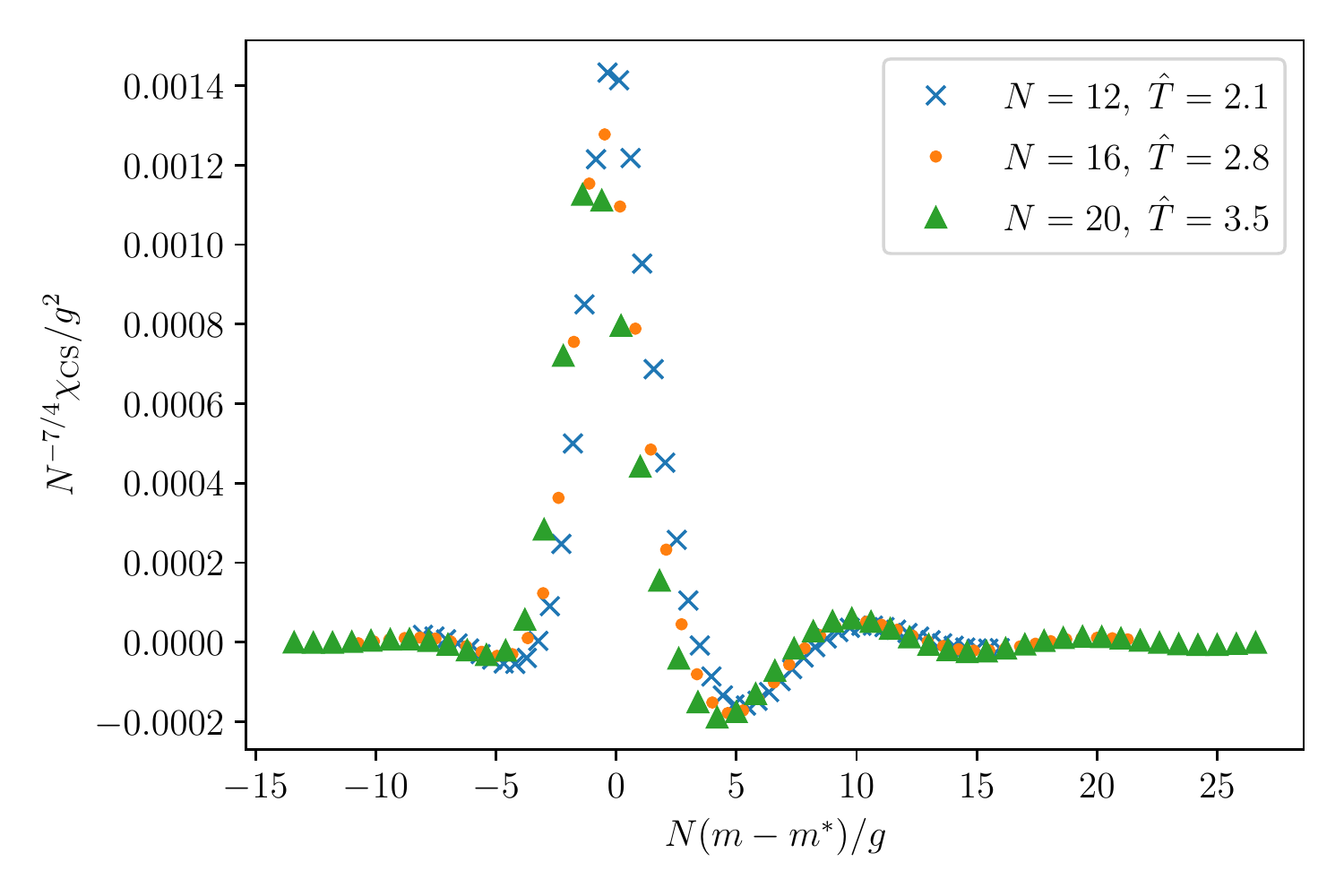}
    \caption{Rescaled dimensionless real-time topological susceptibility $N^{-7/4}\chi/g^2$ as functions of $N(m-m^\ast)/g$ at different values of lattice size $N$. The temporal range $\hat{T}=(ag^2/2)T$ is taken to be proportional to $N$. The topological susceptibility is computed using the Hamiltonian~\eqref{eq:lattice_total_ham3}, where $\theta$ is limited to $0$ or $\pi$.}
    \label{fig:collapse}
\end{figure*}

We have numerically computed the susceptibility (\ref{eq:diffusion}) using a Python package QuSpin~\cite{weinberg2017quspin,weinberg2019quspin}. For the purpose of numerical analysis we convert the lattice Hamiltonian of the Schwinger model to the spin Hamiltonian via the Jordan-Wigner transformation (Appendix~\ref{app:spinHam}). The same Hamiltonian can be directly used to implement the digital quantum simulation of the Schwinger model. 

The real-time topological susceptibilities are shown in Fig.~\ref{fig:TS} as functions of $m/g$ for different values of time $\hat{T}$ and lattice size $N$.
The data exhibits a sharp peak around the critical masses $m^\ast/g\approx-0.33$.
In order to confirm the critical behavior we plot in Fig.~\ref{fig:collapse} the rescaled susceptibility
\begin{align}
    \quad N^{-7/4}\chi_\text{CS}/g^2,
\end{align}
as a function of $N(m-m^\ast)/g$, based on the finite-size scaling analysis detailed in Appendix~\ref{app:scaling}.
We set $m^\ast/g=-0.33$ in accord with previous studies~\cite{schiller1983massive,hamer1982massive,byrnes2002density}. Here, the temporal integral range $\hat{T}$ is taken to be proportional to the spatial lattice size $N$.




The dependence of the susceptibility on the lattice size can be understood using the finite-size scaling, with the critical exponents of the transverse Ising model. The $Z_2$ symmetry of lattice Schwinger model at $\theta=\pi$ and broken parity put it in the same universality class as the $1+1$ dimensional transverse-field Ising model, which in turn is equivalent to 2D Euclidean classical Ising model. The corresponding finite-size scaling analysis is explained in the Appendix~\ref{app:scaling}.


The sharp peak in the real-time topological susceptibility near the critical point of the phase diagram may have important implications for the search for the critical point of the QCD phase diagram. 
It has been argued \cite{halasz1998phase,rajagopal2001condensed} that the behavior near the critical point in the QCD phase diagram belongs to the universality class of 3D Ising model. The model that we have studied in the paper belongs to the universality class of 2D classical Ising model that is characterized by different critical exponents, but shares many common features with the 3D Ising model. 

In particular, we expect that the sharp peak in real-time susceptibility that we have observed near the critical point will also be present in the 3D Ising model, and thus near the critical point in QCD phase diagram. This would imply strong fluctuations of topological charge near the critical point, that can be detected through the charged hadron asymmetries ~\cite{kharzeev2006parity,kharzeev2008effects,kharzeev2016chiral}  that are currently under intense studies at RHIC.

The critical behavior in the model that we have studied belongs to the universality class of model A, in the Hohenberg-Halperin classification \cite{RevModPhys.49.435}. Therefore, we expect that our findings apply to a broad class of physical systems, including those described by the kinetic Ising models \cite{kawasaki1972kinetics}. These models are widely used to describe the relaxational processes in near-equilibrium states, and our study of the real-time critical fluctuations contributes to the field in two different ways.

First, we show how to use the Hamiltonian formalism of quantum field theory in this problem. This formalism is suitable for digital quantum simulations, and opens the pathway towards the study of these non-equilibrium phenomena on future quantum computers. It would be particularly beneficial for the future studies of higher dimensional or more complicated theories such as QCD, where real-time simulations on classical computers are very challenging. 

Second, we demonstrate that close to the critical point the topological susceptibility exhibits a sharp peak. In our case, the topological susceptibility describes the fluctuations of Chern-Simons number, and in other models in the same universality class, such as anisotropic magnets, it would describe the magnetic susceptibility.

The similarity to ferroelectrics may allow one to simulate the real-time ferroelectric response near the critical point~\cite{rowley2014ferroelectric}. Given the importance of ferroelectrics for information storage and processing \cite{dawber2005physics}, this may have interesting practical applications.

\vskip0.3cm

\if{
\section{OTOC in Schwinger model}
\subsection{Spin Hamiltonian}
We study OTOC in massive Schwinger model. According to Kogut-Susskind protocol \cite{PhysRevD.11.395}, the spin Hamiltonian of the massive Schwinger model can be written as  
\begin{equation}
\begin{split}
    H=&\frac{1}{4a}\sum_{n=1}^{N-1}(X_nX_{n+1}+Y_nY_{n+1})+\frac{m}{2}\sum_{n=1}^N(-1)^nZ_n\\
    &+\frac{ag^2}{2}\sum_{n=1}^N\left(\sum_{i=1}^n\frac{Z_i+(-1)^i}{2}+\frac{\theta}{2\pi}\right)^2,
\end{split}
\end{equation}
where $m$ is the fermion mass. The spacial direction is discretized as $x = na$ with an integer $n$ and a lattice constant $a$. The last term is the electric field
\begin{equation}
    L_n=\sum_{i=1}^n\frac{Z_i+(-1)^i}{2}
\end{equation}
and obeys the Gauss law
\begin{equation}
L_{n+1}-L_n=\frac{Z_{n+1}+(-1)^{n+1}}{2}    
\end{equation}
We will drop a background electric field $L_0 = \epsilon_0$, since it plays the same role of $\theta/2\pi$. $\ket{\uparrow\downarrow\uparrow\downarrow\cdots}$ is the ground state in the positive mass limit $m\to\infty$ and $\ket{\downarrow\uparrow\downarrow\uparrow\cdots}$ is the ground state in the negative mass limit $m\to-\infty$. So this model should undergo a phase transition at some $m$.

\subsection{One point functions} 
 Fig. \ref{fig:OPT_N7} and \ref{fig:OPT_N8} present behavior of one-point function $\bra{\Omega}V\ket{\Omega}$ and its square $\bra{\Omega}V^2\ket{\Omega}$ in equilibrium, namely $\ket{\Omega}$ is the respective ground state. They show some abrupt increase around the critical point as indicated by \cite{Kokail19}. In the rest of this note, we set $ \omega=\frac{1}{2a}=0.5,a=g=1$. 

\begin{figure}[H]
    \centering
    \includegraphics[width=\hsize]{OPT.pdf}
    \caption{One point function in equilibrium with [local] $VL_i$ and [average] $V\frac{1}{N}\sum_{n=1}^NL_n$}
    \label{fig:OPT_N7}
\end{figure}

Fig.\ref{fig:Vt} show the mass-dependence of $\bra{\Omega}V(t)\ket{\Omega}$ after a quench. 

\begin{figure}[H]
    \subfigure[]{%
        \includegraphics[clip, width=0.5\columnwidth]{OPT_g1.pdf}}%
    \subfigure[]{%
        \includegraphics[clip, width=0.5\columnwidth]{OPT_g4.pdf}}%
    \caption{$\bra{\Omega}V(t)\ket{\Omega}$ for $N=7$, where $V=\frac{1}{N-1}\sum_{n=1}^NL_n$ and $\Omega=\ket{\downarrow\uparrow\downarrow\uparrow\downarrow\uparrow\uparrow\uparrow}$}
    \label{fig:Vt}
\end{figure}

\subsection{Two point functions}

\begin{figure}[H]
    \centering
    \includegraphics[width=\hsize]{PTs_N7.pdf}
    \caption{$\bra{\Omega} V^2\ket{\Omega}$ in equilibrium with [local] $V=W=L_i$ and [average] $V=W=\frac{1}{N}\sum_{n=1}^NL_n$}
    \label{fig:OPT_N8}
\end{figure}

\begin{figure}[H]
\subfigure[]{%
        \includegraphics[clip, width=0.5\columnwidth]{TPF_g4_without_integrate.pdf}}%
\subfigure[]{%
        \includegraphics[clip, width=0.5\columnwidth]{TPF_N8_g4_without_integrate.pdf}}%
    \caption{$\bra{\Omega}V(t)W\ket{\Omega}$ after quench with $\ket{\downarrow\uparrow\downarrow\uparrow\uparrow\uparrow\downarrow}$ for N=7 and $\ket{\downarrow\uparrow\downarrow\uparrow\downarrow\uparrow\downarrow\uparrow}$ for N=8. $V=W=\frac{1}{N}\sum_{n=1}^NL_n$.}
    \label{fig:my_label}
\end{figure}

 We also consider the time-average of a two point function
\begin{equation}
\label{eq:TPFs}
    \frac{1}{T}\int_0^T \bra{\Omega}W(t)V\ket{\Omega} dt. 
\end{equation}
Figures below are the $N=7$ and $8$ cases. 

\begin{figure}[H]
    \subfigure[]{%
        \includegraphics[clip, width=0.33\columnwidth]{TPF_quench_N7_g_T1.pdf}}%
    \subfigure[]{%
        \includegraphics[clip, width=0.33\columnwidth]{TPF_quench_N7_g.pdf}}%
    \subfigure[]{%
        \includegraphics[clip, width=0.33\columnwidth]{TPF_quench_N7_g_T100.pdf}}%
    \caption{Two point function after quench with $\ket{\Omega}=\ket{\downarrow\uparrow\downarrow\uparrow\uparrow\uparrow\downarrow}$.  $V=W=\frac{1}{N}\sum_{n=1}^NL_n$}
    \label{fig:my_label}
\end{figure}

\begin{figure}[H]
    \subfigure[]{%
        \includegraphics[clip, width=0.5\columnwidth]{TPF_quench_N7_g_T1_28.pdf}}%
    \subfigure[]{%
        \includegraphics[clip, width=0.5\columnwidth]{TPF_quench_N7_g_T20_28.pdf}}%
    \caption{Two point function after quench with $\ket{\Omega}=\ket{\downarrow\uparrow\downarrow\uparrow\downarrow\uparrow\uparrow}$.  $V=W=\frac{1}{N}\sum_{n=1}^NL_n$}
    \label{fig:my_label}
\end{figure}

\begin{figure}[H]
    \subfigure[]{%
        \includegraphics[clip, width=0.5\columnwidth]{TPF_quench_N8_g_T1.pdf}}%
    \subfigure[]{%
        \includegraphics[clip, width=0.5\columnwidth]{TPF_quench_N8_g_T20.pdf}}%
    \caption{Two point function after quench with $\ket{\Omega}=\ket{\downarrow\uparrow\downarrow\uparrow\downarrow\uparrow\downarrow}$.  $V=W=\frac{1}{N}\sum_{n=1}^NL_n$}
    \label{fig:my_label}
\end{figure}

We next consider $\bra{\Omega} V(t)V\ket{\Omega}-\bra{\Omega} V(t)\ket{\Omega}\bra{\Omega}V\ket{\Omega} $. 

\begin{figure}[H]
    \subfigure[]{%
        \includegraphics[clip, width=0.5\columnwidth]{TPF_OPT_g1.pdf}}%
    \subfigure[]{%
        \includegraphics[clip, width=0.5\columnwidth]{TPF_OPT_g4.pdf}}%
    \caption{$\bra{\Omega} V(T)V\ket{\Omega}-\bra{\Omega} V(T)\ket{\Omega}\bra{\Omega}V\ket{\Omega} $ for $N=7$, where $V=\frac{1}{N-1}\sum_{n=1}^NL_n$ and $\Omega=\ket{\downarrow\uparrow\downarrow\uparrow\downarrow\uparrow\uparrow\uparrow}$}
    \label{fig:Vt}
\end{figure}

\begin{figure}[H]
    \subfigure[]{%
        \includegraphics[clip, width=0.5\columnwidth]{TPF_OPT_g1_T100.pdf}}%
    \subfigure[]{%
        \includegraphics[clip, width=0.5\columnwidth]{TPF_OPT_g4_T100.pdf}}%
    \caption{$\bra{\Omega} V(T)V\ket{\Omega}-\bra{\Omega} V\ket{\Omega}\bra{\Omega}V\ket{\Omega} $ for $N=7$, where $V=\frac{1}{N-1}\sum_{n=1}^NL_n$ and $\Omega=\ket{\downarrow\uparrow\downarrow\uparrow\downarrow\uparrow\uparrow}$}
    \label{fig:Vt}
\end{figure}

\begin{figure}[H]
    \subfigure[]{%
        \includegraphics[clip, width=0.5\columnwidth]{TPT_OPT_g1_T30_timeaverage.pdf}}%
    \subfigure[]{%
        \includegraphics[clip, width=0.5\columnwidth]{TPT_OPT_g4_T10_timeaverage.pdf"}}%
    \caption{Time average of $\bra{\Omega} V(t)V\ket{\Omega}-\bra{\Omega} V(t)\ket{\Omega}\bra{\Omega}V\ket{\Omega} $ for $N=7$, where $V=\frac{1}{N-1}\sum_{n=1}^NL_n$ and $\Omega=\ket{\downarrow\uparrow\downarrow\uparrow\downarrow\uparrow\uparrow}$}
    \label{fig:Vt}
\end{figure}

\begin{figure}[H]
    \subfigure[]{%
        \includegraphics[clip, width=0.5\columnwidth]{TPT_OPT_g1_T5_timeaverage.pdf}}%
    \subfigure[]{%
        \includegraphics[clip, width=0.5\columnwidth]{TPT_OPT_g4_T5_timeaverage.pdf}}%
    \caption{Time average of $\bra{\Omega} V(t)V\ket{\Omega}-\bra{\Omega} V\ket{\Omega}\bra{\Omega}V\ket{\Omega} $ for $N=7$, where $V=\frac{1}{N-1}\sum_{n=1}^NL_n$ and $\Omega=\ket{\downarrow\uparrow\downarrow\uparrow\downarrow\uparrow\uparrow}$}
    \label{fig:Vt}
\end{figure}

\subsection{OTOC}
Let $W,V$ be unitary operators. In what follow, we address the OTOC 
\begin{equation}
    \bra{\Omega} W(t)VW(t)V\ket{\Omega}
\end{equation}
where $W(t)=e^{+itH}We^{-itH}$ is the time evolution of $W$ and the expectation value is taken by any state. To diagnose phase transitions, we define the time-average of the OTOC \cite{Heyl_2018}
\begin{equation}
    \frac{1}{T}\int_0^T \bra{\Omega} W(t)VW(t)V\ket{\Omega} dt.
\end{equation}
We use the local electric field $L_i$ for $V$ and $W$.

\begin{figure}[H]
    \centering
    \includegraphics[width=\hsize]{N7_OTOC_without_normalize.pdf}
    \caption{OTOCs in equilibrium with [local] $V=W=L_i$ and [average] $V=W=\frac{1}{N}\sum_{n=1}^NL_n$}
    \label{fig:my_label}
\end{figure}

\begin{figure}[H]
    \centering
    \includegraphics[width=\hsize]{N7_OTOC.pdf}
    \caption{Normalized OTOCs in equilibrium with [local] $V=W=L_i$ and [average] $V=W=\frac{1}{N}\sum_{n=1}^NL_n$}
    \label{fig:my_label}
\end{figure}

We investigate the OTOC after quantum quench, where the expectation value is taken with a fixed state. We use $\ket{\Omega}=\ket{\downarrow\uparrow\downarrow\uparrow\uparrow\uparrow\downarrow}$ and $\ket{\downarrow\uparrow\downarrow\uparrow\downarrow\uparrow\uparrow }$ for $N=7$ and $\ket{\downarrow\uparrow\downarrow\uparrow\downarrow\uparrow\downarrow\uparrow}$ for $N=8$.

\begin{figure}[H]
\subfigure[]{%
        \includegraphics[clip, width=0.33\columnwidth]{quench_OTOC_N7_g_T1.pdf}}%
\subfigure[]{%
        \includegraphics[clip, width=0.33\columnwidth]{quench_OTOC_N7_g_T5.pdf}}%
\subfigure[]{%
        \includegraphics[clip, width=0.33\columnwidth]{quench_OTOC_N7_g_T20.pdf}}%
    \caption{OTOC after quench with $\ket{\downarrow\uparrow\downarrow\uparrow\uparrow\uparrow\downarrow}$.  $V=W=\frac{1}{N}\sum_{n=1}^NL_n$.}
    \label{fig:my_label}
\end{figure}

\begin{figure}[H]
\subfigure[]{%
        \includegraphics[clip, width=0.33\columnwidth]{quench_OTOC_N7_g_T1_28.pdf}}%
\subfigure[]{%
        \includegraphics[clip, width=0.33\columnwidth]{quench_OTOC_N7_g_T5_28.pdf}}%
\subfigure[]{%
        \includegraphics[clip, width=0.33\columnwidth]{quench_OTOC_N7_g_T20_28.pdf}}%
    \caption{OTOC after quench with $\ket{\downarrow\uparrow\downarrow\uparrow\downarrow\uparrow\uparrow}$.  $V=W=\frac{1}{N}\sum_{n=1}^NL_n$.}
    \label{fig:my_label}
\end{figure}

\begin{figure}[H]
\subfigure[]{%
        \includegraphics[clip, width=0.5\columnwidth]{quench_OTOC_N8_g_T1.pdf}}%
\subfigure[]{%
        \includegraphics[clip, width=0.5\columnwidth]{quench_OTOC_N8_g.pdf}}%
    \caption{OTOC after quench with $\ket{\downarrow\uparrow\downarrow\uparrow\downarrow\uparrow\downarrow}$. $V=W=\frac{1}{N}\sum_{n=1}^NL_n$.}
    \label{fig:my_label}
\end{figure}
}\fi

\section*{Acknowledgement}

Y.K. thanks Y.~Meurice for useful comments on the finite-size scaling analysis. This work was supported by the U.S. Department of Energy under awards DE-SC0012704 (D. K. and Y. K.) and DE-FG88ER40388 (D. K.). The work on numerical simulations was supported by the U.S. Department of Energy, Office of Science National Quantum Information Science Research Centers under the ``Co-design Center for Quantum Advantage" award.

\appendix
\section{Massive Schwinger model at finite $\theta$}
\label{app:theta}

We review the massive Schwinger model at finite $\theta$ (see e.g. \cite{tong2018gauge} for a comprehensive review).
Let us start with 1+1D Maxwell theory at finite $\theta$:
\begin{align}
    S = \int\diff^2x\left(\frac{1}{2}E^2 + \frac{g\theta}{2\pi}E\right),
\end{align}
where $E=F_{01}=\p_0A_1=\dot{A}_1$ is an electric field and the temporal gauge $A_0=0$ is taken.
The momentum $\Pi$ conjugate to $A_1$ is,
\begin{align}
 \Pi = \dot{A}_1 + \frac{g\theta}{2\pi},
\end{align}
obeying $[\Pi(x),A_1(y)]=\im\delta(x-y)$.
Hence, the Hamiltonian density takes the form,
\begin{align}
 \calH = \Pi\dot{A} -\left(\frac{1}{2}E^2 + \frac{g\theta}{2\pi}E\right)
 = \frac{1}{2}\left(\Pi-\frac{g\theta}{2\pi}\right)^2.
\end{align}
The Gauss operator must vanish,
\begin{align}
    \p_1\Pi = 0,
\end{align}
on physical states. Therefore, $\Pi$ is fixed once its value is specified, $\Pi|_\text{boundary}=\epsilon_0$, at one of the two boundaries.
The boundary value $\epsilon_0$ is interpreted as a background electric field. Hereafter, we take $\epsilon_0=0$ with the background electric field being absorbed into $\theta$.

\vspace{3mm}

We now add a dynamical Dirac fermion $\psi$ to have,
\begin{align}
\label{eq:cont_Schwinger}
    S = \int\diff^2x\left(\frac{1}{2}E^2 + \frac{g\theta}{2\pi}E\right)
    + \bar{\psi}(\im\slashed{D}-m)\psi.
\end{align}
Then, the Gauss law fixes the electric field in the bulk as
\begin{align}
    \Pi = g\int_\mathrm{boundary}^x\diff y\ \psi^\dag\psi(y),
\end{align}
where $\epsilon_0=0$ is taken. We call $\Pi$ the dynamical electric field because it is induced by fermionic excitations as opposed to $E=\dot{A_1}=\Pi - g\theta/(2\pi)$, where the second term represents the background electric field.

\vspace{3mm}

In line with the main text, we will briefly describe the bosonized form of the theory.
The bosonization dictionary is as follows:
\begin{align}
\begin{split}
    \im\bar{\psi}\gamma^\mu\p_\mu\psi\quad &\leftrightarrow \quad \frac{(\p_\mu\varphi)^2}{2},
    \\
    \bar{\psi}\gamma^\mu\psi\quad &\leftrightarrow \quad -\frac{\epsilon^{\mu\nu}\p_\nu\varphi}{\sqrt{\pi}},
    \\
    \bar{\psi}\gamma^\mu\gamma_5\psi\quad &\leftrightarrow \quad -\frac{\p_\mu\varphi}{\sqrt{\pi}},
    \\
    \bar{\psi}\psi\quad &\leftrightarrow \quad -\frac{cg}{\sqrt{\pi}}\cos(2\sqrt{\pi}\varphi),
\end{split}
\end{align}
with a dimensionless constant $c=\e^\gamma/(2\pi)$ and the Euler constant $\gamma$.
Therefore, the action~\eqref{eq:cont_Schwinger} is converted to
\begin{align}
\begin{split}
    S &= \int\diff^2x\Big[\frac{1}{2}E^2 + \frac{g\theta}{2\pi}E
    \\
    &+ \frac{(\p_\mu\varphi)^2}{2}
    +gA_\mu\frac{\epsilon^{\mu\nu}\p_\nu\varphi}{\sqrt{\pi}}+\frac{cmg}{\sqrt{\pi}}\cos(2\sqrt{\pi}\varphi)\Big]
    \\
    &= \int\diff^2x\Big[\frac{1}{2}E^2 + \frac{g}{2\pi}(2\sqrt{\pi}\varphi+\theta)E
    \\
    &+ \frac{(\p_\mu\varphi)^2}{2} +\frac{cmg}{\sqrt{\pi}}\cos(2\sqrt{\pi}\varphi)\Big].
\end{split}
\end{align}
The equation of motion for $\varphi$ yields,
\begin{align}
    \p^2\varphi = \frac{g}{\sqrt{\pi}}E - 2cmg\sin(2\sqrt{\pi}\varphi),
\end{align}
which, at $m=0$, turns out to be the anomaly equation for the axial current $j^\mu_5=\bar{\psi}\gamma^\mu\gamma_5\psi$.
The equation of motion for $A_1$ yields,
\begin{align}
    \p_1\left(E + g\frac{2\sqrt{\pi}\varphi+\theta}{2\pi}\right)=0.
\end{align}
By requiring the quantity inside the parenthesis to vanish at spatial infinity, we find
\begin{align}
    E = - g\frac{2\sqrt{\pi}\varphi+\theta}{2\pi}.
\end{align}
Using the equation of motion, or equivalently integrating out the gauge field, we get the pure bosonic action,
\begin{align}
\begin{split}
    S = \int\diff^2x\Big[
     &\frac{(\p_\mu\varphi)^2}{2}
     -\frac{g^2}{8\pi^2}(2\sqrt{\pi}\varphi+\theta)^2
     \\
     &+\frac{cmg}{\sqrt{\pi}}\cos(2\sqrt{\pi}\varphi)\Big].
\end{split}
\end{align}
For $4\pi m > g/(\sqrt{\pi}c)$, at $\theta=\pi$ there exist the degenerate ground states for $\varphi$ satisfying,
\begin{align}
    \frac{cmg}{\sqrt{\pi}}\sin(2\sqrt{\pi}\varphi) = \frac{g^2}{4\pi^2}(2\sqrt{\pi}\varphi+\pi).
\end{align}
Given the two solutions $\varphi = (-\pi \pm \calE)/(2\sqrt{\pi})$ with some value $\calE$, we find the associated total electric fields
\begin{align}
    E = \pm g\frac{\calE}{2\pi},
\end{align}
respectively.
For $m\gg g$, $\calE$ approaches $\pi$ and thus the total electric field $E$ approximately takes the values $\pm g/2$.

\section{Spin Hamiltonian of the Schwinger model}
\label{app:spinHam}

We show the spin Hamiltonin used for the numerical simulation presented in the main text and for the purpose of implementing digital quantum simulation. We employ the Jordan-Wigner transformation~\cite{Jordan:1928wi}
\begin{align}
\begin{split}
 \chi_n &= \frac{X_n-\im Y_n}{2}\prod_{i=1}^{n-1}(-\im Z_i),
 \\
 \chi^\dag_n &= \frac{X_n+\im Y_n}{2}\prod_{i=1}^{n-1}(\im Z_i),
\end{split}
\end{align}
to convert the Hamiltonian of the massive Schwinger model to the following spin Hamiltonian~\cite{PhysRevD.11.395},
\begin{equation}
\begin{split}
     H=&\frac{1}{4a}\sum_{n=1}^{N-1}(X_n X_{n+1}+Y_n Y_{n+1})+\frac{m}{2}\sum_{n=1}^N(-1)^n Z_n\\
     &+\frac{a g^2}{2}\sum_{n=1}^N\left(\sum_{i=1}^n\frac{Z_i+(-1)^i}{2}\right)^2.
 \end{split}
 \end{equation}
The last term is the gauge kinetic term with the electric field given by
\begin{equation}
     L_n=\sum_{i=1}^n\frac{Z_i+(-1)^i}{2},
\end{equation}
which is uniquely fixed by the Gauss law
\begin{equation}
L_{n+1}-L_n=\frac{Z_{n+1}+(-1)^{n+1}}{2}    
\end{equation}
upon fixing the boundary electric field to $0$.
 

\section{Finite-volume analysis}
\label{app:scaling}

We are interested in computing the static susceptibility in the Schwinger model,
\begin{align}
\chi(m):=\lim_{\omega\to 0}\lim_{k\to0}\chi(k,\omega,m),
\end{align}
where $\chi(k,\omega;m)$ is the retarded Green's function of electric fields.
In 1+1 dimensional systems, the electric field can be regarded as a topological charge density $E = F_{01}$.

In the thermodynamic limit the Green's function is expected to diverge at the quantum critical point. It is, however, not divergent in finite volume with the system size $N$. Hence, we attempt to identify the critical points, specified by the fermion mass $m^\ast$, by fitting the regularized Green's function $\tilde{\chi}(m;N)$.

Around the critical point $\Delta m := m-m^\ast\sim 0$, the susceptibility behaves as~\cite{Hamer:1981qt},
\begin{align}
\label{eq:inf_chi}
 \chi(m) \approx A|\Delta m|^{-\gamma}.
\end{align}
In a finite system of size $L=Na$, the susceptibility is
\begin{align}
\label{eq:finite_chi}
 \tilde{\chi}(m;N) \approx N^{\gamma/\nu}Q(N/\xi),
\end{align}
with the scaling function $Q$ and the correlation length $\xi$ given by
\begin{align}
\label{eq:xi}
 \xi(m) \approx \xi_0|\Delta m|^{-\nu}.
\end{align}
The ansatz~\eqref{eq:finite_chi} is satisfied in the limit $N\to\infty$ by requiring
\begin{align}
 Q(x) \approx A(\xi_0x)^{-\gamma/\nu} \qquad x\to\infty.
\end{align}
Note the scaling relation can be read off from the relation between the susceptibility and correlation length~(Sec.4 in \cite{sachdev2007quantum}),
\begin{align}
 \chi \sim \xi^{2-\eta}.
\end{align}
Combined with \eqref{eq:inf_chi} and \eqref{eq:xi} we find the scaling relation
\begin{align}
 \gamma = \nu(2-\eta).
\end{align}

The criticality of lattice Schwinger model belongs to the same universality class as 1+1D transverse Ising model, which in turn equivalent to 2D classical Ising model. Hence the critical exponents are,
\begin{align}
 \gamma = \frac{7}{4}, \qquad
 \eta = \frac{1}{4}, \qquad
 \nu = 1.
\end{align}

Finally, the relation~\eqref{eq:finite_chi} in a finite system becomes $N$-independent if we use the following rescaled quantities:
\begin{align}
    N^{1/\nu}\Delta m = N\Delta m,
    \quad
    N^{-\gamma/\nu}\tilde{\chi} = N^{-7/4}\tilde{\chi}.
\end{align}
Indeed, the data points with different lattice size $N$ approximately collapse as shown in Fig.~\ref{fig:collapse}. The susceptibility $\tilde{\chi}$ corresponds to $\chi_\text{CS}$ in the main text.



\bibliographystyle{utphys}
\bibliography{ref}
\end{document}